\documentclass[12pt,fleqn]{article}
\usepackage{amsfonts}
\usepackage{amsmath}
\usepackage{amssymb}
\usepackage{a4wide}
\usepackage{lscape}
\usepackage{mathpazo}
\usepackage{rotating}
\usepackage{subcaption}
\usepackage{float}
\usepackage{epstopdf}
\usepackage{rotating}
\usepackage{adjustbox}
\usepackage{setspace}
\usepackage[margin=0.85in,paper=letterpaper]{geometry}
\usepackage[flushleft]{threeparttable}
\usepackage{booktabs}
\usepackage{adjustbox}
\usepackage{dsfont}
\usepackage{bm}
\usepackage{tikz}
\def\*#1{\mathbf{#1}}
\def\+#1{\boldsymbol{#1}}


\usepackage[english]{babel}
\usepackage[round]{natbib}
\begin{document}

\title{\textsc{Simple Difference-in-Differences Estimation in Fixed-$T$ Panels}\thanks{Westerlund would like to thank the Knut and Alice Wallenberg Foundation for financial support through a Wallenberg Academy Fellowship.}}
\author{\and Nicholas Brown\\
{\small Queen's University} \and Kyle Butts\\
{\small University of Colorado Boulder} \and Joakim Westerlund\thanks{Corresponding author: Department of Economics, Lund University, Box 7082, 220 07 Lund, Sweden. Telephone: +46 46 222 8997. Fax: +46 46 222 4613. E-mail address: \texttt{joakim.westerlund@nek.lu.se}.}\\
{\small Lund University}\\
{\small and}\\
{\small Deakin University}
\date{May 24, 2023}}

\maketitle

\begin{abstract}
\setlength{\baselineskip}{0.83cm}
The present paper proposes a new treatment effects estimator that is valid when the number of time periods is small, and the parallel trends condition holds conditional on covariates and unobserved heterogeneity in the form of interactive fixed effects. The estimator also allow the control variables to be affected by treatment and it enables estimation of the resulting indirect effect on the outcome variable. The asymptotic properties of the estimator are established and their accuracy in small samples is investigated using Monte Carlo simulations. The empirical usefulness of the estimator is illustrated using as an example the effect of increased trade competition on firm markups in China.
\end{abstract}

\setlength{\baselineskip}{0.83cm}



\section{Introduction}

A key assumption in treatment effects studies is that there cannot be any unobserved systematic differences between treated and untreated cross-sectional units in absence of treatment. This is the so-called ``parallel trend'' assumption, which has long been acknowledged to be controversial in practice. Yet there have been surprisingly few formal attempts to resolve the issue, despite the huge empirical literature that has emerged. The standard approach in the panel data context is to assume that any non-parallel trending can be captured using fixed effects. But then this assumption is known to be restrictive. Interactive effects can be used to allow for more general types of non-parallel trending. Here the time effects, or ``common factors'', represent common trends and the individual effects, or `` factor loadings'', measure the extent to which the impact of these trends is equal, or parallel, across units.

\citet{chan2022pcdid} allow for non-parallel trending in the form of interactive effects that are dealt with using a version of the principal components-based approach of \citet{bai2009panel}. However, this method requires that the number of time periods, $T$, is large, and in treatment effects studies $T$ is often small (see \citealp{Bertrand_etal_2004}, for a survey). The approach also requires solving a non-convex optimization problem, which means that it is not only computationally costly but it can also be difficult to get to converge, and even if it does converge it may not be to the global optimum (see \citealp{Moon_Weidner_2019}). \citet{Callaway_Karami_2020} and \citet{brown2022generalized} provide treatment effects estimators that are valid even if $T$ is small. However, these estimators are based on generalized method of moments (GMM), which is computationally burdensome and rely on the availability of certain external instruments. Both estimators require that the number of unobserved factors is know, which is of course never the case in practice.

In this paper, we propose a new treatment effects estimator that is not only valid when $T$ is fixed and the number of factors is unknown but that is also extremely simple to implement. Moreover, unlike most existing estimators, the new estimator is applicable even if the covariates are affected by the treatment status, which is likely to be the case in practice (see \citealp{Caetano_Callaway_Payne_Rodrigues_2022}, for a discussion). It is therefore very attractive from an empirical point of view. This attractiveness is achieved by our novel use of the common correlated effects (CCE) approach of \citet{pesaran2006estimation}, which has a closed form, does not require $T$ to be large and is valid provided only that the number of factors is not larger than the number of observables. The object of interest is the average treatment effect on the treated (ATT), which is the average difference between the actual and counterfactual post-treatment outcomes of treated cross-section units. This average could be computed had it not been for the fact that the counterfactual outcome is unobserved. We therefore have to estimate, or ``impute'', it and this is where the CCE approach come in. The proposed CCE-based difference-in-differences (DD) estimator, dubbed ``C$^2$ED$^2$'' and pronounced ``Cetoo-E-Detoo'', is computed in four steps.\footnote{The name and its pronunciation are inspired by the Star Wars robot character R2-D2.} We begin by estimating the common factors using cross-sectional averages of the outcome variable and covariates from the never-treated sample, as prescribed by CCE. We then estimate the slope coefficients of the controls along with the heterogeneous factor loadings conditional on the first-step factor estimates. In the third step, we use the first- and second-step estimates to estimate untreated covariates in post-treatment periods. In the fourth and final step, we use the first- and second-step estimates together with the third-step estimated covariates to estimate counterfactual outcomes. The estimated ATT is the average difference between the observed treated and estimated counterfactual outcomes.

The new estimator is shown to be consistent and asymptotically normal under very general condition provided only that the number of cross-sectional units, $N$, is large enough, a results that is verified in finite samples by means of a small-scale Monte Carlo simulation study. As an empirical illustration, we consider as an example the effect of increased trade competition on the dispersion of markups in China.

The rest of the paper is structured as follows. Section 2 presents the model and defines the ATT, the estimation of which is the concern of Section 3. Sections 4, 5 and 6 contain the asymptotic, Monte Carlo and empirical studies, respectively. Section 7 concludes. All proofs are relegated to the online appendix.

\section{The model}

We are interested in estimating the ATT of a particular treatment on some outcome variable $y_{i,t}$, observable for $i=1,...,N$ cross-sectional units and $t=1,...,T$ time periods. We allow for the possibility that the $N$ units can be divided into groups within which treatment timing is the same. We follow \citet{Callaway_SantAnna_2020} in defining a treatment group by the time period in which they enter treatment. There are $G$ such groups indexed by $g\in \mathcal{G} \subset \{2,...,T\}$, which for notational convenience is also the period at which the units of group $g$ enter treatment. Hence, if $\mathcal{G} = \{4,8\}$, then there are $|\mathcal{G}| = 2$ groups, the first (second) of which enter treatment in time period $g = t=4$ ($g = t=8$). Treated units never leave their groups but remain exposed for all periods after entering treatment; that is, treatment is of the ``absorbing state''. A unit that is never treated is a member of group $g=\infty$. Treatment timing is randomly assigned conditional on the unobserved interactive effects. Let us therefore denote by $g_i \in \mathcal{G}^+ = \mathcal{G} \cup \{\infty\}$ a random variable stating the group membership of cross-sectional unit $i$, and by $\mathcal{I}_g = \{i: g_i = g \in \mathcal{G}^+\} \subset \{1,...,N\}$ the set of cross-sectional units that are members of group $g$. The set of non-treated units is therefore denoted $\mathcal{I}_\infty$, and it is convenient to let $\mathcal{I}_\infty^c = \{1,...,N\}\backslash \mathcal{I}_\infty$ denote the set of treated units. The number of cross-sectional unit within group $g$ is given by $|\mathcal{I}_g|$. The start of the first treatment is henceforth denoted $g_{\min} = \min\{g_1,...,g_N\}$.

Following the previous literature, we denote by $y_{i,t}(g)$ the ``potential'' outcome of cross-sectional unit $i$ in period $t$ when member of group $g \in \mathcal{G}^+$. Of course, we do not observe $y_{i,t}(g)$ simultaneously for all $g$; instead we observe $y_{i,t} = y_{i,t}(g_i)$, the realized outcome for unit $i$ at time $t$. We may also observe covariates, whose outcome may again depend on treatment status. In our empirical application, the outcome variable is industry-level markup dispersion, treatment is China's ascension into WTO, and a key control variable is the dispersion in marginal-cost. Our analysis allows treatment to affect the dispersion of both prices and marginal-cost and quantify the effect of markup-dispersion on the outcome.

Let us therefore introduce the $m\times 1$ vector $\*x_{i,t}(g)$, whose realized value is given by $\*x_{i,t} = \*x_{i,t}(g_i)$. The model for $y_{i,t}(\infty)$ that we will be considering is given by
\begin{align}
y_{i,t}(\infty) = \+\beta_i'\*x_{i,t}(\infty) +  \+\alpha_i'\*f_t + \varepsilon_{i,t},\label{y0}
\end{align}
where $\+\beta_i$ is a $m\times 1$ vector of heterogeneous slope coefficients, $\*f_t$ is a $r \times 1$ vector of unobservable common factors, $\+\alpha_i$ is a $r\times 1$ vector of factor loadings, and $\varepsilon_{i,t}$ is an idiosyncratic error term.\footnote{The presence of $\+\beta_i'\*x_{i,t}(\infty)$ in \eqref{y0} is an allowance and not a requirement. If there are no regressors, we define $\+\beta_i'\*x_{i,t}(\infty) = 0$. It is important to note, though, that if there are no regressors, the number of factors can be at most one unless there are outside factor proxies ($r \leq 1$), as will be made clear in Section 3.} The interactive effects are given here by $\+\alpha_i'\*f_t$. The purpose of these is to capture non-parallel trending behaviour, that is, unobserved differences in trends between treated and untreated units in absence of treatment. In this terminology, the factors represent common trends and the loadings measure the extent to which the effect of these trends are equal, or parallel, across units. We are not interested in inference on these effects.\footnote{In fact, inference on $\+\alpha_i$ and $\*f_t$ is not even possible, as they are not separately identifiable.} Accurate estimation of $\+\alpha_i$ is therefore not needed.

Unlike $\+\alpha_i$, $\+\beta_i$ is often of some interest. However, since in the present paper $T$ is fixed, we cannot estimate each individual slope accurately. The best that we can hope for is accurate estimation of $\+\beta = \mathbb{E}(\+\beta_i)$. In fact, in many applications in economics (and elsewhere) we are not particularly interested in the marginal effect for a particular unit anyway and so we focus instead on the average marginal effect. The C$^2$ED$^2$ approach enables inference on $\+\beta$ but the main object of interest is as already pointed out the ATT. 

We want to entertain the possibility that $\*x_{i,t}(\infty)$ load on $\*f_t$, because otherwise the factors can be ignored without cost.\footnote{If $\*x_{i,t}(\infty)$ does not load on $\*f_{t}$, $\+\beta_i$ can be estimated by ordinary least squares (OLS) as in \citet{Wooldridge_2005}.} Also, many variables are affected by common shocks, and it is not difficult to find empirical evidence in support of this (see, for example, \citealp{westerlund2019cce}). Let us therefore assume that
\begin{align}
\*x_{i,t}(\infty) = \+\lambda_i'\*f_t + \*v_{i,t}, \label{x}
\end{align}
where $\+\lambda_i$ is a $r \times m$ matrix of factor loadings and $\*v_{i,t}$ is a $m \times 1$ vector of idiosyncratic errors.

We are now ready to introduce the ATT. The treatment effect for unit $i$ at time $t$ when treated in time $g$ is given by
\begin{align}
\Delta_{i,g,t} = y_{i,t}(g) - y_{i,t}(\infty) , \label{te}
\end{align}
Because we do not observe $y_{i,t}(g)$ and $y_{i,t}(\infty)$ simultaneously, $\Delta_{i,g,t}$ must be treated as unknown and estimated from the data. This brings us back to the discussion in the previous paragraph about $\+\beta_i$; because $T$ is fixed, the best that we can hope for is accurate estimation of the ATT, which is the average $\Delta_{i,g,t}$ for group $g$;
\begin{align}
\mathbb{E}(\Delta_{i,g,t}| g_i = g ) = \Delta_{g,t} \label{att}
\end{align}
for $t \geq g \in \mathcal{G}$. Note that while there cannot be any systematic variation across units within groups, we do allow $\Delta_{g,t}$ to vary freely over time and across groups, which means that the effect of the treatment need not take place abruptly at time $g$ but can be gradual in nature. The effect cannot take place prior to treatment, though, which is the so-called ``no anticipation'' condition. Formally, we require that $y_{i,t}(g) = y_{i,t}(\infty)$ for all not-yet-treated observations $t < g\in \mathcal{G}$.\footnote{If treated units anticipate treatment up to $s$ periods before $g$, shift treatment timing to $g - s$.}

Most studies assume that the covariates are unaffected by the treatment and in this case the model for $y_{i,t}(g)$ can be obtained by simply inserting \eqref{y0} into \eqref{te} (see, for example, \citealp{chan2022pcdid}). In the present paper, however, there is no such assumption. In order to be able to separate the part of the ATT that is due to the covariates from the part that is not, we define $\+\tau_{i,g,t} = \*x_{i,t}(g) - \*x_{i,t}(\infty)$ and $\eta_{i,g,t} =  \Delta_{i,g,t} - \+\tau_{i,g,t}'\+\beta_i$. In the terminology of the mediation literature (see, for example, \citealp{huber2014identifying}), $\eta_{i,g,t}$ is the ``direct'' effect of treatment and $\+\tau_{i,g,t}'\+\beta_i$ is the mediated effect of treatment through the covariates, henceforth referred to as the ``indirect'' effect. Hence, provided that $\+\tau_{i,g,t}$ and $\+\beta_i$ are independent, defining $\eta_{g,t} = \mathbb{E}(\eta_{i,g,t} | g_i = g)$ and $\+\tau_{g,t} = \mathbb{E}(\+\tau_{i,g,t} | g_i = g)$, the total ATT can be decomposed as follows:
\begin{align}
\Delta_{g,t} = \eta_{g,t} + \+\tau_{g,t}' \+\beta,
\end{align}
where $\eta_{g,t}$ and $\+\tau_{g,t}' \+\beta$ are the direct and indirect ATTs, respectively.

\section{The C$^2$ED$^2$ estimator}

\subsection{The total ATT}

The estimation of the ATT is carried out using a version of what \citet{borusyak2021revisiting} refer to as the ``imputation'' approach, or what \citet{Xu_2017} refer to as the ``generalized synthetic control'' method, which is based on replacing all unknowns in the definition of $\Delta_{g,t}$ in \eqref{att} by estimates. Note first that since $y_{i,t}(g)$ is observed for treated units in post-treatment periods, we have $y_{i,t} = y_{i,t}(g)$ for treated units post-treatment. Let us therefore turn to $y_{i,t}(\infty)$. We need to estimate this counterfactual for all treated units in post-treatment periods. CCE takes cross-sectional averages of the outcome and covariates as estimators of (the space spanned by) the factors. We tailor this procedure to the present treatment effect scenario where treatment status can affect both outcomes and covariates in unspecified ways. We use never-treated observations to estimate the factors. Then, for the treated units, we estimate the never-treated potential covariates, which are in turn used to estimate the never-treated potential outcomes. This method is detailed in the following four-step procedure to the estimation of $y_{i,t}(\infty)$.

\bigskip

\noindent \textbf{Counterfactual estimation procedure:}

\begin{enumerate}
\item Compute
\begin{align}
\widehat{\*f}_t = \frac{1}{|\mathcal{I}_\infty|}\sum_{i \in \mathcal{I}_\infty} \*z_{i,t}  \label{fhat}
\end{align}
for all $t$, where $\*z_{i,t} = [y_{i,t},\*x_{i,t}']'$ is a $(m+1)\times 1$ vector containing all the observables. The above is the regular CCE estimator of $\*f_t$ computed using the never-treated units only. The fact that $\widehat{\*f}_t$ is computed based on the never-treated units only is crucial since in the present paper both $y_{i,t}$ and $\*x_{i,t}$ may depend on the treatment, and this in turn may well render CCE inconsistent. Equally important is the fact that $\widehat{\*f}_t$ is computed for all time periods $t$. In step 2 the pre-treatment estimates are used to estimate $\+\beta$ and $\{\+\alpha_i\}_{i=1}^N$, while in steps 3 and 4 the post-treatment estimates are used to impute $y_{i,t}(\infty)$ and $\*x_{i,t}(\infty)$ in treatment periods.

\item Estimate the following regression by ordinary least squares (OLS) for all $i$ and $t < g_{\min}$, where $g_{\min}$ again marks the start of the first treatment:
\begin{align}
y_{i,t} = \+\beta'\*x_{i,t} + \*a_i'\widehat{\*f}_t + u_{i,t}. \label{preregr}
\end{align}
Also, $\*a_i$ is a $(m+1)\times 1$ vector of factor loadings and $u_{i,t} = \+\alpha_i'\*f_t - \*a_i'\widehat{\*f}_t + (\+\beta_i-\+\beta)'\*x_{i,t} +  \varepsilon_{i,t}$ is a composite error term. The above OLS regression with $\widehat{\*f}_t$ in place of $\*f_t$ is regular CCE based on the full pre-treatment sample but where $\widehat{\*f}_t$ comes from the subsample of untreated units.\footnote{Note that unlike when using the principal components method, in CCE there is no need to recompute $\widehat{\*f}_t$ if the time period changes, and hence $\{\widehat{\*f}_t\}_{t \geq g_{\min}}$ can be taken directly from step 1.} Define the $(g_{\min}-1)\times 1$ vector $\*y_{i} = [y_{i,1},...,y_{i,g_{\min}-1}]'$, and the $(g_{\min}-1)\times m$ matrices $\*x_{i} = [\*x_{i,1},...,\*x_{i,g_{\min}-1}]'$ and $\widehat{\*f} = [\widehat{\*f}_{1},...,\widehat{\*f}_{g_{\min}-1}]'$. Let $\*M_{\*A} = \*I_{g_{\min}-1} - \*A(\*A'\*A)^{-1}\*A'$ for any $(g_{\min}-1)$-rowed matrix $\*A$. In this notation, the CCE estimators of $\+\beta$ and $\*a_i$ in \eqref{preregr} are given by
\begin{align}
\widehat{\+\beta} &= \left(\sum_{i=1}^N  \*x_{i}'\*M_{\widehat{\*f}}\*x_{i}\right)^{-1}\sum_{i=1}^N \*x_{i}' \*M_{\widehat{\*f}}\*y_{i},\\
\widehat{\*a}_i &= (\widehat{\*f}'\widehat{\*f})^{-1}\widehat{\*f}'(\*y_{i}-\*x_{i}\widehat{\+\beta}),
\end{align}
where the latter estimator is computed for all $i$. The fact that $\widehat{\*a}_i$ is computed for all $i$ is again important, because in step 3, $y_{i,t}(\infty)$ and $\*x_{i,t}(\infty)$ will be estimated for treated units.

\item Compute
\begin{equation}
    \widehat{\*x}_{i,t}(\infty) = \widehat{\+\lambda}_i'\widehat{\*f}_t
\end{equation}
for all treated observations $i\in \mathcal{I}_\infty^c$ and $t \geq g_i$. Here, $\{\widehat{\*f}_t\}_{t \geq g_{\min}}$ is from step 1 and
\begin{equation}
\widehat{\+\lambda}_i = ( \widehat{\*f}' \widehat{\*f} )^{-1} \widehat{\*f}' \*x_i,
\end{equation}
where $\widehat{\*f}$ and $\*x_i$ are the same as in step 2. Note that $\widehat{\+\lambda}_i$ is the OLS estimator of $\+\lambda_i$ in the following regression, which is estimated for each $i\in \mathcal{I}_\infty^c$ individually and $t < g_{\min}$:
\begin{align}
\*x_{i,t} = \+\lambda_i'\widehat{\*f}_t + \*w_{i,t},
\end{align}
where $\*w_{i,t} = \+\lambda_i'(\*f_t - \widehat{\*f}_t)  +  \*v_{i,t}$.

\item The sought counterfactual estimator is given by
\begin{align}
\widehat y_{i,t}(\infty) = \widehat{\+\beta}'\widehat{\*x}_{i,t}(\infty) + \widehat{\*a}_i'\widehat{\*f}_t
\end{align}
which is again available for all treated observations. Here $\widehat{\+\beta}$ and $\{\widehat{\*a}_i\}_{i\in \mathcal{I}_\infty^c}$ are from step 2, $\{\widehat{\*f}_t\}_{t \geq g_{\min}}$ is from step 1, and $\{\widehat{\*x}_{i,t}(\infty)\}_{i\in \mathcal{I}_\infty^c, t \geq g_i}$ comes from step 3.
\end{enumerate}

A few remarks are in order. First, while $\widehat{\+\beta}$ is consistent, $\widehat{\*a}_i$ is not and in fact remains random even asymptotically because $T$ is fixed. Moreover, the asymptotic distribution is not centered at $\+\alpha_i$ but at a certain rotation of $\*a_i$. Interestingly, as we show in Section 3.2, these problems do not interfere with the consistency and asymptotic normality of the estimated ATT.

Second, one can allow $\+\beta$ to vary systematically across groups without affecting the asymptotic validity of the estimated ATT. The only change needed is that the step-2 estimation of this coefficient has to be carried out group-wise, as opposed to just once for all $N$ units. This gives $\{\widehat{\+\beta}_g\}_{g\in \mathcal{G}}$, which should then be inserted instead of $\widehat{\+\beta}$ in step 3.

Third, as \citet{Caetano_Callaway_Payne_Rodrigues_2022} point out, the validity of estimates of the ATT depends on whether or not the covariates are affected by treatment status. For example, if we are estimating the effect of a certain policy aimed at reducing unemployment, we might want to control for the rate of poverty. But then such policies might indirectly reduce poverty, which means that the poverty rate covariate will absorb some of the treatment effect. This is what \citet{angrist2009mostly} call a ``bad control''. It creates a dilemma where including the covariate induces ``post-treatment bias'' and excluding it induces ``omitted variables bias'' (see \citealp{aklin2017can}). In this paper we follow \citet{Caetano_Callaway_Payne_Rodrigues_2022}, and solve this dilemma by imputing and controlling for untreated potential covariates, $\*x_{i,t}(\infty)$. In fact, we go a step further and allow for inference in this indirect effect. 

With $y_{i,t}(g)$ known and $y_{i,t}(\infty)$ estimated, the estimated treatment effect is given by
\begin{align}
\widehat \Delta_{i,g,t} = y_{i,t} - \widehat y_{i,t}(\infty)  \label{theat}
\end{align}
for $i \in \mathcal{I}_g \subset \mathcal{I}_\infty^c$. The estimated ATT for group $g$ at time $t$ is obtained by averaging over the relevant treated group;
\begin{align}
\widehat \Delta_{g,t} = \frac{1}{|\mathcal{I}_g|}\sum_{i \in \mathcal{I}_g} \widehat \Delta_{i,g,t}. \label{atthat}
\end{align}
This is the C$^2$ED$^2$ estimator of $\Delta_{g,t}$.

It is important to note that the C$^2$ED$^2$ estimator does not involve any estimation of the number of factors, $r$. This is in stark contrast to existing principal components-based approaches such those of \citet{chan2022pcdid}, and \citet{Xu_2017}, and GMM approaches such as those of \citet{Callaway_Karami_2020}, and \citet{brown2022generalized}, where asymptotic theory is based on treating $r$ as known. This means that in empirical work, $r$ has to be replaced by an estimator, and accurate estimation of this object is known to be a difficult (see, for example, \citealp{moon2015linear}, and \citealp{breitung2021alternative}). The fact that the proposed estimator does not require estimation of $r$ is therefore a great advantage in practice.

Asymptotic standard errors of estimates of the ATT are generally difficult to compute. Many studies therefore resort to bootstrap inference (see, for example, \citealp{Callaway_Karami_2020}, and \citealp{Xu_2017}), which can be computationally unattractive. We instead employ a version of the non-parametric variance estimator considered by \citet{pesaran2006estimation}. The appropriate estimator to use in our case is
\begin{align}
\widehat \sigma^2(\widehat \Delta_{g,t}) = \frac{1}{|\mathcal{I}_g|-1}\sum_{i \in \mathcal{I}_g} (\widehat \Delta_{i,g,t}  - \widehat \Delta_{g,t})^2.\label{nonparametric variance estimator}
\end{align}
In addition to being simple to compute, non-parametric standard errors are robust and they tend to perform well in small samples (see, for example, \citealp{chudik2011weak}, \citealp{pesaran2006estimation}, and \citealp{westerlund2022cce}).

\subsection{The direct and indirect ATTs}

We demonstrate in Section 2 how the total ATT $\Delta_{i,g}$ can be decomposed into the direct ATT, $\eta_{i,g}$, and the indirect ATT, $\+\tau_{i,g}'\+\beta$. We now demonstrate how to estimate these constituent parts.

The estimator of $\+\tau_{g,t}$ is completely analogous to that of $\Delta_{g,t}$, and is given by
\begin{equation}
\widehat{\+\tau}_{g,t} = \frac{1}{|\mathcal{I}_g|} \sum_{i \in \mathcal{I}_g} \widehat{\+\tau}_{i,g,t}, \label{x treatment effect}
\end{equation}
where $\widehat{\+\tau}_{i,g,t} = \*x_{i,t} - \widehat{\*x}_{i,t}(\infty)$. In the empirical literature, significant estimates of $\widehat{\+\tau}_{g,t}$ is sometimes taken as evidence of indirect treatment effects. However, even if the covariates are affected by treatment, this does not necessarily imply that the outcome is affected, as the effect of changing the covariates on the outcome is determined by their partial effects, here represented by $\+\beta_i$. The proposed C$^2$ED$^2$ approach recognizes this possibility. Our estimate of the indirect ATT is given by the product $\widehat{\+\tau}_{g,t}' \widehat{\+\beta}$, where $\widehat{\+\beta}$ is from step 2 of the counterfactual estimation procedure. Given $\widehat{\+\tau}_{g,t}' \widehat{\+\beta}$, the estimated direct ATT is given by
\begin{equation}
\widehat{\eta}_{g,t} = \widehat{\Delta}_{g,t} - \widehat{\+\tau}_{g,t}' \widehat{\+\beta}.
\end{equation}

The variances of $\widehat{\+\tau}_{g,t}$ and $\widehat{\eta}_{g,t}$ can be estimated non-parametrically in the following obvious way:
\begin{align}
\widehat{\+\Sigma}(\widehat{\+\tau}_{g,t}) & = \frac{1}{|\mathcal{I}_g| - 1} \sum_{i \in \mathcal{I}_g} ( \widehat{\+\tau}_{i,g,t} - \widehat{\+\tau}_{g,t} ) ( \widehat{\+\tau}_{i,g,t} - \widehat{\+\tau}_{g,t} )' ,\\
\widehat{\sigma}^2(\widehat{\eta}_{g,t}) & = \frac{1}{|\mathcal{I}_g| - 1} \sum_{i \in \mathcal{I}_g} ( \widehat{\eta}_{i,g,t} - \widehat{\eta}_{g,t} )^2.
\end{align}
Note that $\widehat{\sigma}^2(\widehat{\eta}_{g,t})$ is a direct estimator of the variance of the estimated direct ATT. The corresponding estimator of the variance of the estimated indirect ATT is given by $\widehat{\+\beta}' \widehat{\+\Sigma}(\widehat{\+\tau}_{g,t}) \widehat{\+\beta}$.

The above estimator of $\eta_{g,t}$ is of the plug-in type; it takes the definition of $\eta_{g,t}$ and plugs in estimates in places of true quantities. An alternative estimation approach is to take $\widehat \Delta_{g,t}$ but to replace $\widehat{\*x}_{i,t}(\infty)$ with $\widehat{\*x}_{i,t}$ when computing $\widehat y_{i,t}(\infty)$ in step 4 of the counterfactual estimation procedure. The fact that changing the way that the covariates enter in step 4 alters the object being estimated is important not only for the present paper but also when considering the works of others. As mentioned earlier, \citet{chan2022pcdid} proposes a principal components-based estimator of the ATT that assumes that the covariates are unaffected by treatment and they use the observed covariates in their estimations. Logic based on our findings suggests that if the unaffected covariates assumption is false, Chan and Kwok's estimator will only capture the direct ATT. In the empirical illustration of Section 6, we elaborate on this point. 

\section{Asymptotic results}

In this section, we study the asymptotic properties of the estimated total ATT and its direct and indirect parts. The conditions that we will be working under are given in Assumptions 1--9. Here and throughout, $\mathrm{tr}\, \*A$, $\mathrm{rank}\, \*A$ and $\|\*A\| = \sqrt{\mathrm{tr}\,(\*A'\*A)}$ denote the trace, the rank, and the Frobenius (Euclidean) norm of the generic matrix $\*A$, respectively. The symbols $\to_d$ and $\to_p$ signify convergence in distribution and probability, respectively.

\bigskip

\noindent \textbf{Assumption 1.} $g_{\min} > m+2$.

\bigskip

\noindent \textbf{Assumption 2.} $\mathrm{plim}_{N\to\infty} |\mathcal{I}_g|/N \in (0,1)$ for all $g \in \mathcal{G}^+$.

\bigskip

Assumptions 1 and 2 are sample size conditions. They ensure that $g_{min}$ is large enough to ensure that the step-2 regression model in \eqref{preregr} is feasible and also that each group is non-negligible as $N$ increases, which is necessary for accurate estimation of the group-specific ATTs. We write Assumption 2 in terms of convergence in probability because $|\mathcal{I}_g|$ is a random quantity.

\bigskip

\noindent \textbf{Assumption 3.} $\+\beta_{i} = \+\beta + \+\nu_i$, $\Delta_{i,g,t}= \Delta_{g,t} + \upsilon_{i,t}$, and $\+\tau_{i,g,t} = \+\tau_{g,t} + \+\zeta_{i,t}$ where $\+\nu_i$, $\upsilon_{i,t}$, and $\+\zeta_{i,t}$ are independently distributed across $i$ and $t$ with zero mean, and finite fourth-order cumulants.

\bigskip

Assumption 3 is a random parameter condition that is largely the same as in \citet{chan2022pcdid}, and \citet{Gobillon_Magnac_2016}. None of parameters are required to be heterogeneous, as the covariance matrices of $\+\nu_i$, $\upsilon_{i,t}$ and $\+\zeta_{i,t}$ need not be positive definite.

Before we continue onto Assumption 4, is it useful to first lay out some additional notation. Step 1 of the counterfactual estimation procedure uses the cross-sectional averages of the observables in $\*z_{i,t}$ for the untreated units to estimate the factors. This means that both $y_{i,t}$ and $\*x_{i,t}$ have to be informative of those factors. By combining (\ref{y0}) and (\ref{x}) we arrive at the following static factor model for $\*z_{i,t}$:
\begin{align}
\*z_{i,t} = \+\Lambda_i'\*f_t + \*e_{i,t}, \label{z}
\end{align}
where $\+\Lambda_i = [\+\alpha_i +\+\lambda_i\+\beta_i, \+\lambda_i]$ is $r\times (m+1)$ and $\*e_{i,t} = [\varepsilon_{i,t} + \+\beta_i'\*v_{i,t}, \*v_{i,t}']'$ is $(m+1)\times 1$. This expression for $\*z_{i,t}$ implies that $\widehat{\*f}_t$ can be written in the following way:
\begin{align}
\widehat{\*f}_t = \frac{1}{|\mathcal{I}_{\infty}|}\sum_{i \in \mathcal{I}_{\infty}} \*z_{i,t} = \frac{1}{|\mathcal{I}_{\infty}|}\sum_{i \in \mathcal{I}_{\infty}} \+\Lambda_i'\*f_t + \frac{1}{|\mathcal{I}_{\infty}|}\sum_{i \in \mathcal{I}_{\infty}} \*e_{i,t}.
\end{align}
Assumptions 4--6 below ensure that the average $\*e_{i,t}$ tends to zero as $N$ increases and that the average $\+\Lambda_i$ has full row rank, which in turn ensure that $\widehat{\*f}_t$ is consistent for the space spanned by $\*f_t$.

\bigskip

\noindent \textbf{Assumption 4.} $\varepsilon_{i,t}$ and $\*v_{i,t}$ are independently distributed across $i$ with zero mean, and finite fourth-order cumulants.

\bigskip

\noindent \textbf{Assumption 5.} $\*f_t$, $g_i$, $\varepsilon_{i,t}$, $\*v_{i,t}$, $\+\nu_i$, $\+\zeta_{i,t}$, and $\upsilon_{i,t}$ are mutually independent.

\bigskip

\noindent \textbf{Assumption 6.} $\mathrm{rank}(|\mathcal{I}_{\infty}|^{-1}\sum_{i \in \mathcal{I}_{\infty}} \+\Lambda_i ) = r \leq m+1$ almost surely.

\bigskip

\noindent \textbf{Assumption 7.} The $r \times r$ matrix $\sum_{t=1}^T\*f_t\*f_t'$ is positive definite for all $T$.

\bigskip

\noindent \textbf{Assumption 8.} $N^{-1}\sum_{i=1}^N \*x_{i}'\*M_{\widehat{\*f}} \*x_{i} \to_p \+\Sigma$ as $N\to\infty$, where the $m\times m$ matrix $\+\Sigma$ is positive definite.

\bigskip

Assumptions 7 and 8 are standard non-collinearity conditions. Assumption 7 generalizes the usual ``within assumption'' in the individual fixed effects only model, which rules out time-invariant regressors. Assumption 7 rules out more general ``low-rank'' regressors, as it is almost always done in models with interactive effects (see \citealp{moon2015linear}, for a discussion). The exclusion restriction is not very restrictive, though, as it does not rule out low rank regressors in the model for $y_{i,t}$. If there are such regressors present, then these should be treated as observed factors, which can be appended to $\widehat{\*f}_t$ in step 1 of the counterfactual estimation procedure, as we illustrate in Section 6. This is an advantage in the sense that while $\+\beta_{i}$ and $\Delta_{i,g,t}$ are subject to the random parameter condition in Assumption 3, $\+\alpha_i$ is not. Hence, unlike the coefficients of the observed covariates, the coefficients of low rank regressors are not restricted in any way. The disadvantage of this observed factor treatment of low rank regressors is that we cannot estimate their coefficients.

An important point about Assumptions 1--8 is that the time series properties of $\*f_t$, $\varepsilon_{i,t}$, $\*v_{i,t}$ and $\Delta_{i,g,t}$ are essentially unrestricted. \citet{chan2022pcdid} allow for non-stationary factors and regressors (in a large-$T$ setting) but the regression errors have to be stationary, which is tantamount to requiring that the observables are cointegrated with the factors. Assumptions 1--8 are more general in this regard. One implication of this generality is that as long as $m+1\geq r$ there is no need to model the deterministic component of the data, as deterministic regressors can be treated as additional (unknown) factors to be estimated from the data. If there are common known deterministic terms, such as an intercept or a linear time trend, these can be inserted into $\widehat{\*f}$ along with the cross-sectional averages. As with the dynamics, the type of heteroskedasticity that can be permitted is not restricted in any way.

We are now ready to state Theorem 1, which contains our two main results.

\bigskip

\noindent \textbf{Theorem 1.} \emph{Under Assumptions 1--8, as $N\to\infty$,}
\begin{itemize}
  \item[(a) ] $\begin{aligned}[t]
    \frac{\sqrt{|\mathcal{I}_g|}(\widehat \Delta_{g,t} - \Delta_{g,t})}{\sigma(\widehat \Delta_{g,t})} \to_d N(0, 1 ),
    \end{aligned}$

  \item[(b) ] $\begin{aligned}[t]
    \widehat \sigma^2(\widehat \Delta_{g,t}) \to_p \sigma^2(\widehat \Delta_{g,t}),
    \end{aligned}$
\end{itemize}
\emph{where the definition of $\sigma^2(\widehat \Delta_{g,t})$ is provided in the appendix.}

\bigskip

The proof of Theorem 1 is contained in the appendix, where we show that $\sqrt{|\mathcal{I}_g|}(\widehat \Delta_{g,t} - \Delta_{g,t})$ is asymptotically mixed normal, and that this implies that $\sqrt{|\mathcal{I}_g|}(\widehat \Delta_{g,t} - \Delta_{g,t})/\sigma(\widehat \Delta_{g,t})$ is asymptotically standard normal. This result is unintuitive given the inconsistency of $\widehat{\*a}_i$ in step 2 of the counterfactual estimation procedure, as mentioned earlier. The reason is that the asymptotic distribution of $\widehat{\*a}_i$ is centered at a rotated version of $\*a_i$, and that the effect of this rotation is absorbed in the estimation of $\*f_t$. The asymptotic distribution of $\widehat \Delta_{i,g,t} - \Delta_{i,g,t}$ is therefore correctly centered at zero despite the inconsistency, and it is independent across $i$. Asymptotic normality is therefore possible after averaging over the relevant subsample.

Another point about Theorem 1 is that it holds even if $r$ is unknown, provided only that $m+1 \geq r$, so that the number of factors is not under-specified. As we show in the proof, while $\sigma^2(\widehat \Delta_{g,t})$ depends on whether $m+1 = r$ or $m+1 > r$, this dependence is successfully mimicked in large samples by $\widehat \sigma^2(\widehat \Delta_{g,t})$. We can therefore show that
\begin{align}
\frac{\sqrt{|\mathcal{I}_g|}(\widehat \Delta_{g,t} - \Delta_{g,t})}{\widehat \sigma^2(\widehat \Delta_{g,t})} = \frac{\sqrt{|\mathcal{I}_g|}(\widehat \Delta_{g,t} - \Delta_{g,t})}{\sigma^2(\widehat \Delta_{g,t})} + o_p(1) \to_d N(0, 1 )
\end{align}
as $N\to\infty$. Asymptotically valid inference is therefore possible for any $r$ satisfying $m+1 \geq r$. This robustness is particularly important given the well-known bias problem of post-selection estimators \citep{leeb2005model}.

The asymptotic distributions of the direct and indirect ATTs are a direct consequence of Theorem 1 and the consistency of $\widehat{\+\beta}$, and are summarized in the following corollary.

\bigskip

\noindent \textbf{Corollary 1.} \emph{Suppose that the conditions of Theorem 1 are met. Then, as $N\to\infty$,}
\begin{itemize}
  \item[(a)] $\begin{aligned}[t]
    \frac{\sqrt{|\mathcal{I}_g|} ( \widehat{\+\tau}_{g,t}'\widehat{\+\beta} - \+\tau_{g,t}'\+\beta )}{\sqrt{\widehat{\+\beta}'\widehat{\+\Sigma}(\widehat{\+\tau}_{g,t})\widehat{\+\beta}}  } \to_d N(0, 1 ),
    \end{aligned}$

  \item[(b)] $\begin{aligned}[t]
    \frac{\sqrt{|\mathcal{I}_g|}(  \widehat \eta_{g,t} - \eta_{g,t} )}{\widehat{\sigma}(\widehat{\eta}_{g,t})}  \to_d N(0,1).
    \end{aligned}$
\end{itemize}

\section{Monte Carlo simulations}

In this section, we present the results of a small-scale Monte Carlo study. The processes used to generate the potential treated outcome and covariates, $y_{i,t}(\infty)$ and $\*x_{i,t}(\infty)$, respectively, are given by restricted versions of \eqref{y0} and \eqref{x} that set $r=m=2$ and $\*f_t = [1,t]'$. Equation \eqref{x} is generated with $\+\lambda_i = \*I_2 + \*Z_i$, where the elements of $\*Z_i$ are drawn independently from $N(0,1)$, as are the elements of $\*v_{i,t}$. Equation \eqref{y0} is generated with $\+\beta_i = \+\beta = [1, 1]'$ for all $i$ and $\+\alpha_i \sim \mathrm{diag}(\+\lambda_i) + \+\theta d_i + N([0,0]', \*I_2)$, where $d_i = 1(i \in \mathcal{I}_\infty^c)$ is a dummy that is one if cross-section unit $i$ is treated and zero otherwise, and $\mathrm{diag}(\+\lambda_i)$ vectorizes the main diagonal of $\+\lambda_i$. The term $\+\theta d_i$ controls whether the parallel trend condition is met. If $\+\theta = [0,0]'$, then $\mathbb{E}(\+\alpha_i) = [1,1]'$ for all $i$ and so trends are on average parallel, whereas if $\+\theta = [0,1]'$, then $\mathbb{E}(\+\alpha_i) = [1,1+d_i]'$, and so the treated and untreated cross-sectional units are on different trend paths. The presence of $\mathrm{diag}(\+\lambda_i)$ makes $\+\alpha_i$ correlated with $\+\lambda_i$, which in turn means that $\*x_{i,t}(g)$ is endogenous. The regression errors are allowed to be serially correlated through $\varepsilon_{i,t} = \rho \varepsilon_{i,t-1} + u_{i,t}$, where $\varepsilon_{i,0} = 0$, $\rho = 0.75$ and $u_{i,t} \sim N(0,1)$.

The potential treated outcome and covariates are generated as $y_{i,t}(g) = \Delta_g + y_{i,t}(\infty)$ and $\*x_{i,t}(g) = \+\tau_g + \*x_{i,t}(\infty)$, respectively, which means that in this data generating process the direct treatment effect is given by $\eta_g =  \Delta_g - \+\tau_g'\+\beta$. We assume that there is just one treated group and randomly assign half of the cross-sectional units to this group. Consistent with the empirical illustration of Section 6, we set $N=164$ and $T=9$. Treatment starts in period seven, and so $g = g_{\min} = 7$. As for $\+\tau_g$ and $\Delta_g$, we consider two cases. In the first, $\Delta_g = 1$ and $\+\tau_g = [0,0]'$, and therefore the direct ATT is given by $\eta_g = \Delta_g = 1$, whereas in the second, $\Delta_g = 2$ and $\+\tau_g = [0,1]'$, which means that while $\eta_g = 1$ is the same as before, now there is also an indirect ATT equal to $\+\tau_g'\+\beta = 1$.

The C$^2$ED$^2$ procedure is implemented exactly as described in Section 3. We focus on the total ATT. The results for the direct and indirect ATTs were very similar and are available upon request. The C$^2$ED$^2$ results are compared to those obtained by using two-way fixed effects OLS with one treatment dummy for each of the three treatment periods, which represents the workhorse of the empirical treatment effects literature. We consider two specifications; one that accounts for the covariates and one that ignores them. For each estimator, we report the average bias and the mean squared error (MSE). The number of replications is set to 1,000.

\begin{center}
{\sc Insert Tables \ref{tab:monte_results_pt} and \ref{tab:monte_results_no_pt} about here}
\end{center}

Tables \ref{tab:monte_results_pt} and \ref{tab:monte_results_no_pt} report the results for the cases when the parallel trend condition holds and when it fails, respectively. We begin by considering Table \ref{tab:monte_results_pt}. Since in this case trends are parallel and the covariates are unaffected by the treatment, even the OLS estimator that omits the covariates is expected to be unbiased, which is just what we see in the table. The ranking of the three estimators in terms of MSE is also as expected with the C$^2$ED$^2$ estimator that accounts for both factors and covariates outperforming the competition. The covariate-augmented OLS estimator is biased when the indirect ATT is nonzero. This is due in part to the correlation between $\+\alpha_i$ and $\+\lambda_i$, which causes an omitted variables bias when the covariates are included but the factors are not appropriately accounted for, in part to the fact that controlling for the covariates absorbs the indirect ATT, as pointed out in Section 3. According to Table \ref{tab:monte_results_no_pt}, if trends are not parallel, OLS breaks down regardless of whether it is covariate-augmented or not, which is again just as expected because fixed effects OLS is inconsistent in this case.

\section{Empirical illustration}

One of the channels through which competition may affect gains from trade is via changes in markups, which measures the ability of firms to charge prices above their marginal costs. As is well-known, first-best efficiency is obtained when markups are the same across goods. Of course, in practice markups are never the same and this raises the possibility of so-called ``pro-competitive'' effects of trade, which is the idea that trade liberalization through increased competition drive down both the level and dispersion of markups, leading to increased efficiency. Moreover, welfare improves when consumers benefit from lower markups of the goods they consume and when producers gain from higher markups in foreign markets.

Since its accession into the World Trade Organization (WTO) in the end of 2001, China's role in the world economy has grown enormously. As a result, the pro-competitive effects of China's WTO accession have attracted considerable attention, so much so that there is by now a separate strand of literature devoted to them. The bulk of the evidence seem to suggest that both the level and dispersion of markups have gone down following the WTO entry, and that this development has had important welfare effects (see, for example, \citealp{Hsu_etal_2020}).

The purpose of the current application is to contribute to the above mentioned literature. This is done in two ways. First, we account for general forms of unobserved heterogeneity. The standard approach in the literature is to exploit differences in tariffs across industries. The basic idea is to split the sample of industries into a treatment and a control group, where the former is assumed to be relatively more exposed to the WTO accession. Given that pre-WTO tariffs varied greatly across industries, the argument goes on to say that industries that had previously been protected with relatively high tariffs experienced greater tariff reduction. They should therefore be relatively more exposed. The effect of the WTO accession is then estimated via a standard DD-style OLS regression in which markup is regressed onto a dummy variable that takes on the value one for treated industries in post-WTO periods, control variables, and industry and time fixed effects.

While popular, the standard approach to WTO evaluation has (at least) two drawbacks. One drawback is that it requires that in absence of treatment the difference between the treatment and control groups is constant over time. Trends therefore have to be parallel, which is known to be restrictive. A very commonly cited reason is that certain industries have more lobbying power for protection. Tariffs may be granted in response to domestic special interest groups, the pressure of which may vary over time (see, for example, \citealp{Fan_etal_2018}, \citealp{Deng_etal_2018}, and \citealp{Xiang_etal_2017}). Differences in lobbying power may therefore cause the treatment and control groups to differ systematically over time even if China had not joined the WTO in 2001.\footnote{Similarly, policymakers may lower tariffs selectively only in industries that are able to compete with relatively less expensive imports, for example, in industries experiencing a productivity boom (see \citealp{Brandt_etal_2017}).} Such differences are problematic as they render the fixed effects OLS estimator inconsistent, as the Monte Carlo results of Section 5 illustrates. The main problem is that many sources of possible non-parallel trending are unknown and lack good proxies. For this reason, in lack of better alternatives, it is common to control for industry-specific linear time trends (see, for example, \citealp{Liu_Qiu_2016}, and \citealp{Mao_Xu_2019}). Deterministic trends can account for some non-parallel trending but not all. Moreover, results tend to be highly sensitive to the inclusion of such trends, which reinforces the sentiment in the literature that non-parallel trending is an important issue.\footnote{Some studies include common controls that are thought to be highly correlated with various kinds of protectionism, such as wage rates, employment, exports, and imports (see, for example, \citealp{Hsu_etal_2020}). Again the results tend to be very sensitive.}

Another drawback of the standard approach is that it is not designed to deal with the case when both the outcome and the covariates are affected by treatment. This is important because the literature has identified many channels through which the WTO accession may affect markups (see \citealp{Mao_Xu_2019}, \citealp{Fan_etal_2018}, \citealp{Deng_etal_2018}, \citealp{Liu_Ma_2021}, and \citealp{Brandt_etal_2017}, to mention a few). Two common examples are the price- and cost-change channels. Markup is defined as the ratio of price to marginal cost. This means that markup changes can emanate from price changes, cost changes, or both. It is therefore common to include one of these variables as a covariate and also to estimate the effect of the WTO accession on them (see, for example, \citealp{Mao_Xu_2019}, \citealp{Fan_etal_2018}, and \citealp{lu2015trade}). But then we know from Section 3 that treatment-affected covariates require special treatment or else the estimated ATT will be misleading. Specifically, the inclusion of such covariates will absorb the indirect ATT. Some researchers seem to be aware of this. The following quotation, taken from \citet[page 116]{Fan_etal_2018}, is quite suggestive: ``If the marginal-cost channel indeed plays a role, then once the marginal costs are included as an explanatory variable, we would witness attenuation of the impact of input tariffs on markups.'' However, it is not until recently that researchers in econometrics have considered the possibility of treatment-affected covariates, and there is still much to do (see \citealp{Caetano_Callaway_Payne_Rodrigues_2022}). Empirical researchers therefore have little or no option but to either ignore the problem or to exclude all potentially bad controls from their specifications.

The present paper is not the first to point to these shortcomings, but it is the first to consider an econometric approach that is designed to deal with both in a rigorous way. The C$^2$ED$^2$ approach allows for interactive effects in which there may be unobserved differences between cross-sectional units that change over time as a result of common shocks. The parallel trend condition is therefore not required, which is a substantial advantage when compared to the standard fixed effects-based approach. Another advantage of the approach that we exploit in this section is that it not only allows for covariates that may be affected by treatment but that it makes it possible to assess the relative importance of the direct and indirect treatment channels. It should therefore be well suited for the problem at hand.

The data set that we use is taken from \citet{lu2015trade} (see also \citealp{Deng_etal_2018}, who use the same data), and comprise 164 industries (three-digit Chinese industrial classification) observed over the 1998--2005 period. The smallness of $T$ here, which is a feature of most data sets in the literature, means that it is important to use techniques that work even if $T$ is not large. The Monte Carlo results reported in Section 5 suggest that the proposed C$^2$ED$^2$ approach should work well. Following \citet{lu2015trade}, the outcome variable is markup dispersion, as measured by the markup Theil index (in logs). Industries are assigned to the treatment and control groups based on whether they faced tariffs above or below the sample median in 2001.

Our preference to focus on the \citet{lu2015trade} study is motivated in part by their analysis of the price- and cost-change channels (see their Section E). As a proxy for marginal costs, the authors use productivity (TFP). The ATT is estimated via an OLS regression that in addition to fixed effects, controls and the treatment variable includes the TFP Theil index as a covariate to account for cost dispersion effects. The authors argue that this should allow them to partially isolate the price-change channel. In order to assess the ATT of the WTO accession on costs, the authors run a second OLS regression with the TFP Theil index as dependent variable and the treatment variable as a covariate. The estimated ATTs are significant, which is taken as evidence to suggest that both channels are operational. The purpose of this illustration is to assess the accuracy of this last conclusion.

The above discussion suggests that in terms of the notation of Section 2, in this section $y_{i,t}$ is the markup Theil index, and $\*x_{i,t}$ is the TFP Theil index. The estimated factors in $\widehat{\*f}_t$ are made up of the cross-sectional averages of these variables. A constant is included as an observed factor (as explained in Section 4), which is tantamount to allowing for industry fixed effects. We therefore allow for one known and two unknown factors.

\begin{figure}
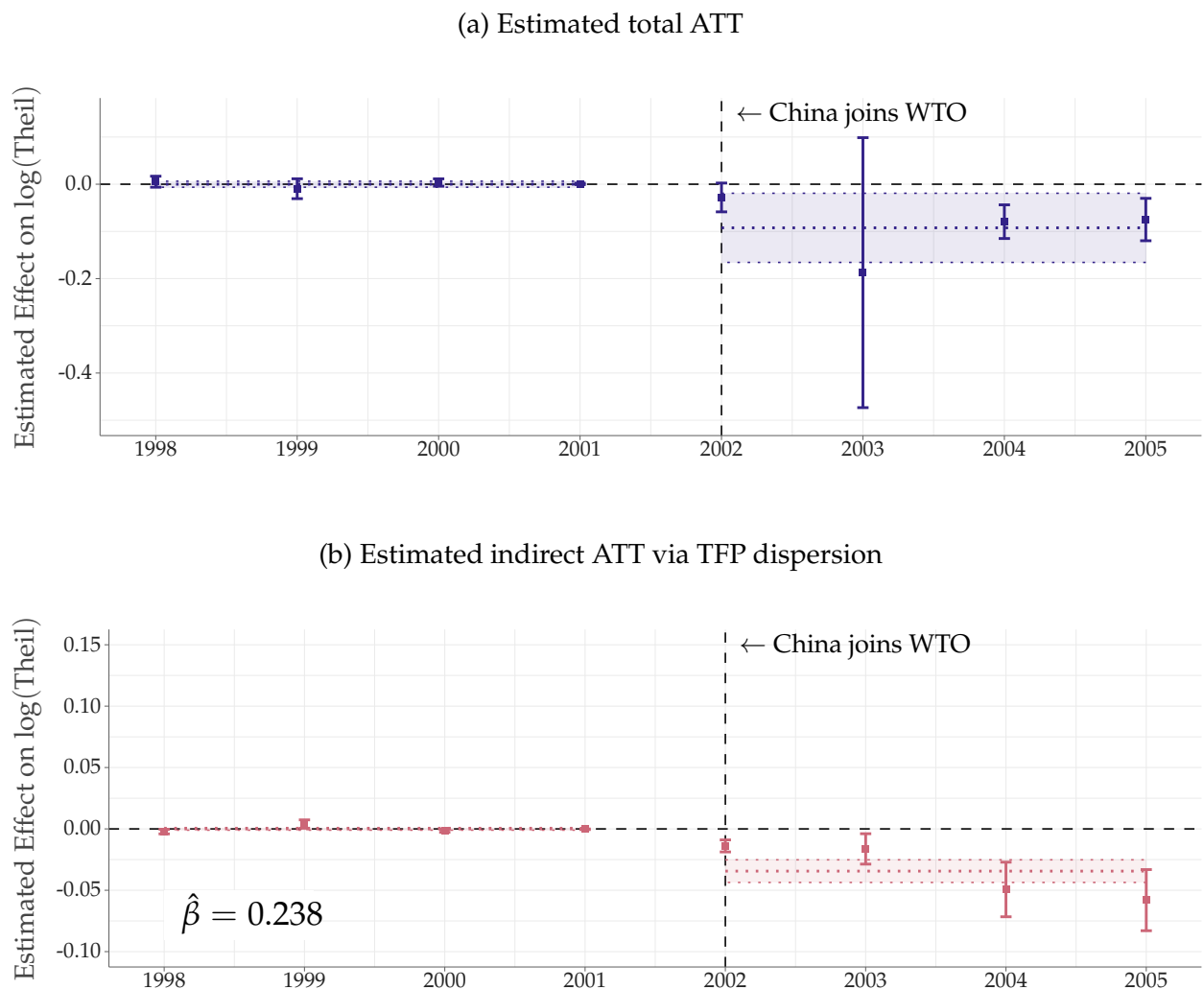

    \caption{Estimated ATTs of China's WTO accession in 2001 on the markup Theil index.}
    \label{fig:trade}

    \begin{subfigure}[b]{\textwidth}
        \caption{Estimated total ATT}
        \input{figures/trade-cce_est.tex}
    \end{subfigure}

    \begin{subfigure}[b]{\textwidth}
        \caption{Estimated indirect ATT via TFP dispersion}
        \input{figures/trade-cce_mediated_est.tex}
    \end{subfigure}

    {\footnotesize\emph{Notes:} The figures present ATT estimates and 95\% confidence intervals for the effect of China's WTO accession in 2001 on the dispersion of markups as measured by the markup Theil index. The treatment group comprise all industries that in 2001 had above-median tariff rates. Estimates are computed using the C$^2$ED$^2$ estimator with the TFP Theil index as a covariate. A constant is included as an observed factor. Figure (a) presents estimates of the total ATT and figure (b) presents the estimated indirect ATT operating through the TFP Theil index. $\hat{\beta}$ in figure (b) refers to the estimated slope on the TFP Theil index in the markup Theil index regression.}
\end{figure}

\begin{center}
{\sc Insert Figure \ref{fig:trade} about here}
\end{center}

The estimated direct and indirect ATTs are reported in Figure \ref{fig:trade}. The estimates are reported for each year and averaged over all the post- and pre-treatment periods, as is customary in the literature. Both types are reported together with 95\% confidence intervals. The first thing to note is that the both the direct and indirect ATTs are estimated to be negative, suggesting that markup dispersion decreased more after 2001 in industries with relatively high tariffs in 2001. Given that industries with higher tariffs in 2001 experienced greater tariff reduction after 2002, these results imply that the WTO accession reduced markup dispersion. We also note that the pre-treatment estimates are all very close to zero, which means that in this period there were no differences in the markup Theil index that depended on group membership.

While insignificant in 2002 and 2003, the year-specific total ATTs reported in Figure \ref{fig:trade} (a) are significant in 2004 and 2005. The point estimate in 2003 is notably noisy. A possible reason for this is that the industry classification system changed in 2003, as noted by, for example, \citet{Chen_etal_2019}, and \citet{lu2015trade}. The estimated average ATT during the whole post-treatment period is about $-0.1$ and significant, which consistent with the results of \citet{Chen_etal_2019}.

In order to assess to what extent the decrease in markup dispersion is due to decreases in TFP dispersion as predicted by the marginal cost channel we look at the estimated indirect ATTs. According to the results reported in Figure \ref{fig:trade} (b), the estimated direct ATTs are negative and significant in the post-treatment period and insignificant in the pre-treatment period. \citet{lu2015trade} estimate the ATT on the TFP Theil index and find it to be significantly negative; however, their approach does not allow them to infer whether this negative response of the TFP Theil index has an effect on the markup Theil index. According to our results, the estimated indirect ATTs are sizable, accounting for almost half of the total ATTs. This is important in itself but also for what it means for the results reported by \citet{lu2015trade}, which are based on including the TFP Theil index as a covariate. In particular, we know from before that this type of conditioning will absorb the indirect effect. In this case, since both ATTs are estimated to be negative and the magnitude of the indirect ATTs are about half of the direct ATTs, conditioning on the TFP Theil index will lead to an underestimation of the total ATTs by about 50\%. This illustrates quite clearly the importance of being able to account for the fact that treatment may affect not only the outcome variable but also the covariates.

\section{Conclusion}

In this paper we propose a new ATT estimator dubbed ``C$^2$ED$^2$'' that is applicable even when the parallel trends condition fails because of the presence of unobserved heterogeneity in the form of interactive fixed effects. Our identification strategy, based on the popular CCE approach, relies on the presence of covariates that load on the same factors as the outcome variable. This allows us to use the cross-sectional averages of the observables to impute the untreated potential outcomes in post-treatment time periods. The covariates are allowed to depend on the treatment status, and if they do C$^2$ED$^2$ makes it possible separate the direct ATT that is unrelated to the covariates from the indirect ATT that works through those covariates. The estimator is shown to be consistent and asymptotically normal, thereby enabling standard inference, provided only that the number of cross-sectional units, $N$, is large, which is a great advantage in practice because in the literature many data sets involve only a few time periods.

\pagebreak

\bibliography{references}

\pagebreak

\begin{table}
\def\arraystretch{1.25}
\caption{Monte Carlo results when trends are parallel.}\label{tab:monte_results_pt}

\begin{center}
\begin{adjustbox}{max width=\textwidth}
\begin{threeparttable}
    \begin{tabular}{@{} l @{\extracolsep{4pt}}cccccc @{}}
    \toprule \addlinespace[3mm]

    & $\text{BIAS}(\widehat{\Delta}_7)$ & $\text{MSE}(\widehat{\Delta}_7)$
    & $\text{BIAS}(\widehat{\Delta}_8)$ & $\text{MSE}(\widehat{\Delta}_8)$
    & $\text{BIAS}(\widehat{\Delta}_9)$ & $\text{MSE}(\widehat{\Delta}_9)$
    \\
    \cmidrule{2-7}

    \multicolumn{7}{@{}l}{
Direct effect only
    } \\
    \midrule \addlinespace[3mm]

OLS & -0.00 & 1.06 & -0.02 & 4.17 & -0.03 & 9.32 \\
OLS with covariates & -0.01 & 0.54 & -0.01 & 1.85 & -0.01 & 3.92 \\
C$^2$ED$^2$ & -0.01 & 0.58 & -0.02 & 1.07 & -0.03 & 1.72 \\

    \multicolumn{7}{@{}l}{
    Direct and indirect effects
    } \\
    \midrule \addlinespace[3mm]

OLS & 0.01 & 1.07 & 0.02 & 4.20 & 0.03 & 9.38 \\
OLS with covariates & -4.19 & 18.21 & -4.28 & 20.31 & -4.35 & 23.07 \\
C$^2$ED$^2$ & -0.02 & 0.57 & -0.03 & 1.07 & -0.04 & 1.69 \\

    \bottomrule
    \end{tabular}

    \begin{tablenotes}[flushleft] \footnotesize
    \item \textit{Notes}: Data are generated for $N=164$ cross-sections and $T=9$ time periods to match the sample used in the empirical illustration. Treatment starts in period $g_{\min} = 7$. $\widehat{\Delta}_7$--$\widehat{\Delta}_9$ are the estimated total ATT for the post-treatment time periods. ``$\text{BIAS}(\widehat{\Delta}_t)$'' and ``$\text{MSE}(\widehat{\Delta}_t)$'' refer to the bias and MSE of the estimated ATT at post-treatment time period $t$, respectively. ``OLS'' and ``OLS with covariates'' refers to the two-way fixed effects OLS estimator without and with covariates, respectively. The results are reported for two data generating processes; one in which there is only a direct effect and one in which there is both direct and indirect effects.
    \end{tablenotes}
\end{threeparttable}
\end{adjustbox}
\end{center}
\end{table}

\begin{table}
\def\arraystretch{1.25}
\caption{Monte Carlo results when trends are not parallel.}\label{tab:monte_results_no_pt}

\begin{center}
\begin{adjustbox}{max width=\textwidth}
\begin{threeparttable}
    \begin{tabular}{@{} l @{\extracolsep{4pt}}cccccc @{}}
    \toprule \addlinespace[3mm]

    & $\text{BIAS}(\widehat{\Delta}_7)$ & $\text{MSE}(\widehat{\Delta}_7)$
    & $\text{BIAS}(\widehat{\Delta}_8)$ & $\text{MSE}(\widehat{\Delta}_8)$
    & $\text{BIAS}(\widehat{\Delta}_9)$ & $\text{MSE}(\widehat{\Delta}_9)$
    \\
    \cmidrule{2-7}

    \multicolumn{7}{@{}l}{
Direct effect only
    } \\
    \midrule \addlinespace[3mm]

OLS & 4.00 & 17.07 & 8.00 & 68.09 & 12.00 & 153.23 \\
OLS with covariates & 4.01 & 16.59 & 8.01 & 66.08 & 12.02 & 148.37 \\
C$^2$ED$^2$ & -0.03 & 1.17 & -0.06 & 2.26 & -0.06 & 3.55 \\

    \multicolumn{7}{@{}l}{
    Direct and indirect effects
    } \\
    \midrule \addlinespace[3mm]

OLS & 4.00 & 17.10 & 8.01 & 68.44 & 12.01 & 153.96 \\
OLS with covariates & -0.19 & 0.68 & 3.71 & 15.76 & 7.63 & 62.27 \\
C$^2$ED$^2$ & -0.06 & 1.20 & -0.06 & 2.36 & -0.06 & 3.64 \\

    \bottomrule
    \end{tabular}

    \begin{tablenotes}[flushleft] \footnotesize
    \item \textit{Notes}: See Table \ref{tab:monte_results_pt} for an explanation.
    \end{tablenotes}
\end{threeparttable}
\end{adjustbox}
\end{center}
\end{table}

\end{document}